\documentstyle{article}
\font\sma=cmr10 at 8truept
\font\big=cmr10 at 12truept
\font\bigbf=cmbx10 at 14truept 

\def\al{\alpha}
\def\be{\beta}
\def\ga{\gamma}
\def\de{\delta}

\def\et{\eta}

\def\la{\lambda}

\def\ch{\chi}
\def\ps{\psi}

\def\Ga{\Gamma}

\def\cl{{\cal L}}
\def\fr#1#2{{{#1} \over {#2}}}
\def\prt{\partial}

\def\pt#1{\phantom{#1}}

\def\half{{\textstyle{1\over 2}}}
\def\frac#1#2{{\textstyle{{#1}\over {#2}}}}

\def\lsim{\mathrel{\rlap{\lower4pt\hbox{\hskip1pt$\sim$}}
    \raise1pt\hbox{$<$}}}
\def\gsim{\mathrel{\rlap{\lower4pt\hbox{\hskip1pt$\sim$}}
    \raise1pt\hbox{$>$}}}
\def\sqr#1#2{{\vcenter{\vbox{\hrule height.#2pt
         \hbox{\vrule width.#2pt height#1pt \kern#1pt
         \vrule width.#2pt}
         \hrule height.#2pt}}}}

\def\cO{{\cal O}}

\newcommand{\beq}{\begin{equation}}
\newcommand{\eeq}{\end{equation}}
\newcommand{\bea}{\begin{eqnarray}}
\newcommand{\eea}{\end{eqnarray}}
\newcommand{\rf}[1]{(\ref{#1})}

\renewenvironment{thebibliography}[1]
 { \rm
   \begin{list}{\arabic{enumi}.}
    {\usecounter{enumi} \setlength{\parsep}{0pt}
     \setlength{\itemsep}{3pt} \settowidth{\labelwidth}{#1.}
     \sloppy
    }}{\end{list}}

\begin{document}

\baselineskip=12pt
{}\pt{x}
\begin{flushright}
{IUHET 431\\}
{February 2001\\}
\end{flushright}
\vglue 46 truept 

\begin{flushright}
{\bigbf TOPICS IN LORENTZ AND CPT VIOLATION\pt{.}%
\vskip -16 pt
\footnote{\sma
Presented at Orbis Scientiae 2000, 
Fort Lauderdale, Florida, December 2000}
\\}
\end{flushright}

\vglue 24 truept 

\begin{flushleft}
{\big V. Alan Kosteleck\'y%
\footnote{\sma
Physics Department, Indiana University, 
Bloomington, IN 47405, U.S.A.}
\\}
\end{flushleft}

\vglue 24 truept 
\rm

\noindent
{\bf 1.\ INTRODUCTION}
\vglue 0.4 cm 

Invariance under Lorentz and CPT transformations
is a fundamental requirement
of local relativistic quantum field theories,
including the standard model of particle physics.
This invariance also seems to be realized in nature,
as no clear signals for violations
have been observed despite numerous experimental tests.
Nonetheless,
several facts offer motivation 
for theoretical studies of possible Lorentz and CPT violation
\cite{cpt98}.
One is that quantitative statements about
the degree to which nature exhibits Lorentz and CPT symmetry 
are best expressed within 
a consistent and general theoretical framework
that allows for violations
\cite{kp,ck,kl}. 
Another more subtle fact is that
the exceptional sensitivity of present experimental tests
\cite{pdg}
implies access to highly suppressed Lorentz and CPT violations
that might arise at scales well beyond the standard model 
in the context of Planck-scale physics
\cite{kps}.

At earlier conferences in this series
\cite{os}
I have discussed the possibility that Lorentz and CPT symmetry
might be violated as a result of new physics 
in a theory underlying the standard model,
perhaps including string theory
\cite{kps}.
I have also described 
the Lorentz- and CPT-violating standard-model extension
that allows for the associated low-energy effects 
in a very general context
\cite{kp,ck}
and outlined some of the numerous existing and future experiments
that test these ideas.
These experiments include,
for example,
studies of neutral-meson oscillations 
\cite{kexpt,kp,ckv,bexpt,ak,ktevproc},
comparative tests of QED 
in Penning traps \cite{bkr,gg,hd,rm},
spectroscopy of hydrogen and antihydrogen \cite{bkr2,dp},
measurements of muon properties
\cite{bkl,vh},
clock-comparison experiments
\cite{ccexpt,kla,lh,db},
observations of the behavior of a spin-polarized torsion pendulum
\cite{bk,bh},
measurements of cosmological birefringence \cite{cfj,ck,jk,pvc},
and observations of the baryon asymmetry \cite{bckp}.

In the present talk I briefly describe 
some of our recent theoretical analyses, 
emphasizing in particular topics
at the level of quantum field theory
\cite{kp,ck}
and the issues of causality and stability
in Lorentz-violating theories 
\cite{kl}.
Other studies of Lorentz and CPT violation in the context of the
standard-model extension
are being presented at this conference
\cite{tw,dc}.
The reader may also find of interest 
recent efforts to create a classical analogue for CPT violation
\cite{jr,kr},
which lie outside the scope of this talk.
A treatment of experimental results released 
in the year since the previous conference in this series
is also outside the present scope.
I mention here only the newly published results 
on Lorentz and CPT violation involving 
protons
\cite{dp}
and neutrons
\cite{db},
and the preliminary announcement of results in the muon sector
obtained from muonium hyperfine spectroscopy 
\cite{vh}.

\vglue 0.6 cm 
\noindent
{\bf 2.\ CONCEPTUAL BASICS}
\vglue 0.4 cm 

Over the past decade,
a framework allowing for Lorentz and CPT violation
within realistic field-theoretic models
has been developed
\cite{kps}
that leads to a phenomenology for Lorentz and CPT violation
at the level of the standard model and quantum electrodynamics (QED)
\cite{kp}.
The resulting general standard-model extension 
\cite{ck}
can be chosen to preserve
the usual SU(3)$\times$SU(2)$\times$U(1) gauge structure
and to be power-counting renormalizable.
Energy and momentum are conserved,
and conventional canonical methods for quantization apply.
In this part of the talk,
I summarize some useful basic concepts
for the theoretical results to follow.

\it Observer and particle Lorentz transformations. \rm
By construction, the standard-model extension
is invariant under rotations or boosts of an observer's inertial frame,
called observer Lorentz transformations.
These must be contrasted with
rotations or boosts of the localized fields 
in a fixed observer coordinate system,
called particle Lorentz transformations, 
which can change the physics
\cite{ck}.
The observer Lorentz invariance 
of the standard-model extension,
together with its generality,
means that it is 
the low-energy limit of \it any \rm realistic underlying theory 
in which the physics is coordinate independent in an inertial frame 
but in which Lorentz symmetry is broken.
Lorentz- and CPT-violating effects could therefore
provide a unique low-energy signature for 
qualitatively new physics from the Planck scale.

One attractive scenario for generating Lorentz- and CPT-violating terms 
while maintaining observer Lorentz invariance
is to invoke spontaneous Lorentz breaking 
in an underlying fully Lorentz-covariant theory at the Planck scale,
perhaps string theory
\cite{kps}.
Since observer Lorentz invariance constrains 
the physical behavior under suitable coordinate changes 
made by an external observer,
an underlying theory incorporating this property
cannot lose it through internal interactions
such as those leading to spontaneous Lorentz violation.
If instead the underlying theory were to
break Lorentz invariance explicitly,
then observer Lorentz invariance would appear unnatural
and imposing it would involve an extra requirement.

\it Stability and causality. \rm
Two crucial features of acceptable physical quantum field theories
are stability and causality.
In relativistic field theories,
these are closely linked to Lorentz invariance
\cite{wp}.
Among their implications 
are the requirements of energy positivity at arbitrary momenta
and of commutativity of spacelike-separated observables,
both of which must hold in all observer inertial frames.
For the standard-model extension,
which allows for (particle) Lorentz violation,
it is natural to ask about the implications of
these requirements 
both within the theory and in the larger context
of the underlying Planck-scale physics
\cite{kl}.

In addressing many aspects of this question, 
the complications of the full standard-model extension
can be avoided by limiting attention 
to the case of the quadratic fermion part of 
a general renormalizable lagrangian 
with explicit Lorentz- and CPT-breaking terms.
This is the single-fermion limit of the free-matter sector
in the general standard-model extension,
given by
\cite{ck}:
\beq
\cl = \half i\overline{\psi} \Ga^\nu
\stackrel{\leftrightarrow}
{\prt}_{\nu}
\hspace{-.1cm}{\psi}
-\overline{\psi}M{\psi},
\label{lagr}
\eeq
where
\beq
{\Ga}^{\nu}\equiv {\ga}^{\nu}+c^{\mu \nu}
{\ga}_{\mu}+d^{\mu \nu}{\ga}_{5} {\ga}_{\mu}
+e^{\nu}+if^{\nu}{\ga}_{5}
+\frac{1}{2}g^{\la \mu \nu}
{\sigma}_{\la \mu}
\label{Gam}
\eeq
and
\beq
M\equiv m+a_{\mu}{\ga}^{\mu}+b_{\mu}{\ga}_{5}
{\ga}^{\mu}+\frac{1}{2}H^{\mu \nu}
{\sigma}_{\mu \nu}.
\label{M}
\eeq
The parameters
$a_{\mu}$, $b_{\mu}$, $c_{\mu\nu}$, $\ldots$, $H_{\mu \nu}$
control the degree of Lorentz and CPT violation,
and their properties are discussed in Ref.\ \cite{ck}.
Note that observer Lorentz symmetry is manifest in this model
because the lagrangian \rf{lagr} 
is independent of the choice of coordinate system.
However,
particle Lorentz transformations modify the fields
but leave invariant the coefficients
$a_{\mu}$, $b_{\mu}$, $\ldots$, $H_{\mu \nu}$,
thereby breaking Lorentz symmetry.

A satisfactory investigation of stability and causality 
for Eq.\ \rf{lagr} or the standard-model extension
must incorporate the implications of observer Lorentz invariance
\cite{kl}.
For example,
if energy positivity is found to be violated 
for specified coefficients for Lorentz violation
in some inertial frame,
then a corresponding difficulty must be present 
in every other inertial frame.
Conversely,
if complete consistency can be established in any inertial frame,
then observer Lorentz invariance ensures consistency
in all other inertial frames.

For simplicity and definiteness in what follows,
I take the mass $m$ of the fermion in Eq.\ \rf{lagr} to be nonzero.
This is certainly appropriate for the non-neutrino fermionic sector 
of the standard-model extension,
and also applies if neutrinos have mass
with possible minor modifications for Majorana fermions.
Many of the results obtained can also be applied 
to bosons and to the massless case,
although some care would be required to handle correctly 
the additional complications arising from
distinctions between finite- and zero-mass
representations of the Lorentz group.
This is particularly true for gauge bosons
\cite{kl},
for which a satisfactory analysis of causality and stability
remains an open problem.
Some results about causality 
restricted to the special case 
of a single Lorentz-violating term 
in the photon sector
have recently been obtained
\cite{fk}.

\it Concordant frames. \rm
The coefficients for Lorentz violation
$a_{\mu}$, $b_{\mu}$, $c_{\mu\nu}$, $\ldots$, $H_{\mu \nu}$
in Eq.\ \rf{lagr} transform as nontrivial representations
of the observer Lorentz group O(3,1).
Since this group is noncompact,
individual components of these coefficients 
can become arbitrarily large.
Under certain circumstances,
including both exact and perturbative calculations,
it is therefore valuable to define a special class of inertial frames
called concordant frames,
in which the coefficients for Lorentz and CPT violation
represent only a small perturbation relative to the ordinary Dirac case
\cite{kl}.

Since no Lorentz and CPT violation has been observed in nature,
any effects are presumably minuscule in an Earth-based laboratory.
Barring unexpected surprises such as fortuitous cancellations,
this suggests that all the coefficients 
in Eq.\ \rf{Gam} are well below 1
and those in Eq.\ \rf{M} are well below $m$. 
It can then be regarded as an experimental fact that
any inertial frame in which the Earth moves nonrelativistically
is a concordant frame.

If small Lorentz- and CPT-violating effects in nature
indeed arise from an underlying theory 
at some large scale $M_P$,
such as the Planck scale,
then the natural dimensionless suppression factor 
is some power of the ratio $m/M_P$
\cite{kp}.
The size of the coefficients
$c_{\mu\nu}$, $d_{\mu\nu}$, $e_{\mu}$, $f_\mu$, $g_{\la\mu\nu}$
in Eq.\ \rf{Gam} is therefore likely to be no larger than $m/M_P$,
although smaller values are possible.
Similarly,
the size of the coefficients
$a_{\mu}$, $b_{\mu}$, $H_{\mu \nu}$
in Eq.\ \rf{M} is likely to be
no larger than $m^2/M_P$.

\it High-energy physics. \rm
Within standard special relativity,
the separation between high- and low-energy physics
is frame independent.
Thus,
high-energy physics in one inertial frame
corresponds to high-energy physics in another frame.
However,
this correspondence fails in the presence of Lorentz violation
\cite{kl}.
The point is that the coefficients for Lorentz and CPT violation
determining the physics of a high-energy particle
in one inertial frame can be very different
from those determining the high-energy physics in another frame.
The breaking of particle Lorentz invariance
therefore implies that high-energy physics 
can change between inertial frames,
despite the observer Lorentz invariance.
In particular, 
statements concerning Lorentz-breaking effects
restricted to high energies may be observer dependent.

Since the standard concept of high and low energy is ambiguous,
a cleaner definition is useful
\cite{kl}.
A useful option is to take
the separation between high and low energies 
relative to the scale of the underlying theory
as being defined in a concordant frame.
This definition is experimentally reasonable
and compatible with intuition and common usage,
since any laboratory frame moves nonrelativistically 
relative to a concordant frame,
so high- and low-energy physics are similar in both. 

\vglue 0.6 cm
\noindent
{\bf 3.\ QUANTUM MECHANICS AND QUANTUM FIELD THEORY}
\vglue 0.4 cm

In a concordant frame,
where the coefficients for Lorentz violation are small,
the usual methods of relativistic quantum mechanics
and quantum field theory can be adopted.
The first step is the construction 
of the relativistic quantum hamiltonian $H$
from the lagrangian $\cl$ of Eq.\ \rf{lagr}.
Care is required because $\cl$ contains extra time-derivative terms.
A spinor redefinition can be used to eliminate
these couplings in a concordant frame
\cite{bkr}:
define $\psi =A\chi$
and require the nonsingular spacetime-independent matrix $A$
to obey $A^{\dagger}\ga^0\Ga^0A=I$.
The reader is warned that 
the explicit form of $A$ depends on the chosen inertial frame.
In any case,
it follows that $\cal L[\chi]$ contains no time derivatives 
other than the usual one,
$\frac{1}{2}{\it i}\overline{\chi}
{\ga}^0\hspace{-.15cm} \stackrel{\leftrightarrow}
{\prt}_0\hspace{-.1cm}{\chi}$.
Since the conversion of $\ps$ to $\ch$
can be regarded as a change of basis in spinor space, 
the physics is unaffected.

It is known that $A$ exists if and only if 
all the eigenvalues of $\ga^0\Ga^0$ are positive.
Quantitatively,
a parameter $\de^0$ can be defined as the upper bound 
on the size of certain coefficients for Lorentz and CPT violation
such that $A$ exists
\cite{kl}.
It can be proved that $\de^0<1/480$,
which represents a value far larger than the 
maximum size of $\de^0$ likely to be acceptable on experimental grounds.
The spinor redefinition involving $A$ therefore 
always exists in a concordant frame
and is applicable to realistic situations in nature. 

After implementing the spinor redefinition from $\ps$ to $\ch$,
one can use the Euler-Lagrange equations to obtain 
a modified Dirac equation in terms of $\ch$.
This takes the form
\beq
(i{\prt}_0-H){\chi}=0 ,
\label{dirac}
\eeq
where the hamiltonian
$H=A^{\dagger}\ga^0(i\Ga^j {\prt}_j-M)A$
is hermitian.
Various explicit forms for this hamiltonian can
be found in Ref.\ \cite{kla}.

The modified Dirac equation \rf{dirac} is solved
via a superposition of plane spinor waves,
as usual:
$\ch (x) = w(\la)\exp(- i \la_\mu x^\mu)$.
The quantity ${\la}_{\mu}$ obeys the dispersion relation
\beq
\det(\Ga^\mu {\la}_{\mu}-M) = 0 .
\label{disp}
\eeq
This dispersion relation is displayed as an explicit polynomial
in Ref.\ \cite{kl}.
It can be regarded as a quartic equation for $\la^0(\vec{\la})$.
For a particle with definite 3-momentum,
the dispersion relation fixes the exact eigenenergies
in the presence of Lorentz and CPT violation.
All four roots of Eq.\ \rf{disp} are necessarily real,
since $H$ is hermitian in a concordant frame.
They are also independent of the spinor redefinition.
The dispersion relation is observer Lorentz invariant,
so $\la_{\mu}$ must be an observer Lorentz 4-vector.

In the usual Dirac case,
the roots of the dispersion relation exhibit a fourfold degeneracy.
However,
in the presence of Lorentz and CPT violation
this degeneracy is typically lifted.
Nonetheless,
for sufficiently small
Lorentz and CPT violation,
the roots can still be separated into two positive ones 
and two negative ones. 
The criterion for the existence of this separation can be 
quantitatively expressed in terms of a parameter $\de$,
defined in Ref.\ \cite{kl}.
It is known that the bound $\delta<m/124$ is sufficient.
The values of this bound is again much larger than
experimental observations are likely to allow,
so the existence of Lorentz and CPT violation in nature
would have no effect on the separation between positive and negative roots.
Note that this bound is independent of the spinor redefinition.
The two bounds on $\de^0$ and $\de$ provide criteria
quantitatively constraining the definition of a concordant frame.

As usual,
the eigenfunctions of the modified Dirac equation \rf{dirac}
corresponding to the two negative roots
can be reinterpreted as positive-energy
reversed-momentum wave functions.
The four resulting spinors $u$ and $v$ are eigenvectors
of the hermitian hamiltonian $H$.
They span the spinor space
and can be used to write a general solution of Eq.\ \rf{dirac}
as a Fourier superposition of plane wave solutions
with complex weights
in the standard way.

To convert from relativistic quantum mechanics to
quantum field theory,
the complex weights in the Fourier expansion 
are promoted to creation and annihilation operators on a Fock space,
as usual.
The spinors $\ps$ and $\ch$ become quantum fields,
related through the redefinition $\ps = A \ch$ as before.
The standard nonvanishing anticommutation relations
can be imposed on the creation and annihilation operators.
The resulting equal-time anticommutators for the fields $\ch$
are conventional,
while the nonvanishing equal-time anticommutators 
for the original fields $\psi$ become
\cite{kl}
\beq
\{\psi_j(t,\vec{x}),
\overline{\psi}_l(t,\vec{x}^{\:\prime})
\Ga^0_{lk}\}
=\delta_{jk}\delta^3(\vec{x}-\vec{x}
^{\:\prime}) ,
\label{fieldcomm}
\eeq
where the spinor indices are explicitly shown. 
Note the generalization from the usual Dirac case
of the canonical conjugate of $\psi$,
which in the presence of Lorentz and CPT violation takes the form
$\pi_{\psi}= \overline{\psi}\Ga^0$,
with $\Ga^0$ given by Eq.\ \rf{Gam}.

In a concordant frame,
the vacuum state $|0\rangle$ of the Hilbert space
is defined as the state that vanishes 
when the annihilation operators are applied.
The creation operators then act on $|0\rangle$
to yield states describing particles and antiparticles
with 4-momenta appropriately determined 
by the dispersion relation \rf{disp}.
An important consequence is 
that the zero components of these 4-vectors are positive definite.
This implies positivity of the energy for the Hilbert-space states
in the concordant frame
\cite{kl}.

\vglue 0.6 cm
\noindent
{\bf 4.\ STABILITY AND CAUSALITY}
\vglue 0.4 cm

The derivation of the quantum physics 
associated with the lagrangian \rf{lagr} can be performed
as outlined in the previous section
provided the constraints on $\de^0$ and $\de$ are satisfied.
These constraints ensure
that the Lorentz-violating time-derivative terms can be eliminated 
and that the usual separation holds between particles and antiparticles,
and they quantify the notion of concordant frame.
They involve specific components of the parameters
for Lorentz and CPT violation
and so are noninvariant under observer Lorentz transformations,
as expected.
A class of observers therefore exists
for whom these bounds are violated and  
the derivation of the quantum physics fails.
These observers are strongly boosted relative to a concordant frame.
Nonetheless,
when combined with the requirement of observer Lorentz invariance,
their existence indicates some difficulty must occur 
even for the quantization scheme in a concordant frame.
These are associated with the
stability and causality of the theory,
which are considered next.

\it Stability. \rm
In conventional relativistic field theory with Lorentz symmetry,
energy positivity in a given frame
implies that the vacuum is stable in any frame
provided certain conditions are met.
One is that for all one-particle states in the given frame
the 4-momenta are timelike or lightlike with nonnegative 
zeroth components.
Since the signs of these zeroth components
are invariant under an observer Lorentz transformation,
energy positivity is a Lorentz-invariant concept 
despite being a statement about a 4-vector component.
Thus,
for example,
the usual free Dirac theory exhibits energy positivity
in all observer frames.

In contrast,
for a theory with Lorentz and CPT violation,
results demonstrating energy positivity in one frame
are insufficient to ensure energy positivity 
or stability in all inertial frames.
This is true,
for example,
of the energy positivity discussed in the previous section
for the theory \rf{lagr} in a concordant frame.
At least one of the usual assumptions fails:
certain energy-momentum 4-vectors
satisfying the dispersion relation \rf{disp}
may in fact be spacelike in all observer frames.
An example is provided by the dispersion relation 
for the special case of the theory \rf{lagr} 
in which only the $b_{\mu}$ coefficient is nonzero.
No matter how small $b_\mu$ is chosen,
an observer frame can always be found
in which spacelike 4-vectors ${\la^{\mu}}$
exist that satisfy the dispersion relation 
\cite{kl}.

Observer Lorentz invariance ensures that the instabilities
due to the existence of spacelike solutions 
exist in any frame, including concordant ones.
However,
they are perhaps most intuitively appreciated 
by considering an appropriate observer boost.
With a boost velocity less than 1,
it is always possible to 
convert a spacelike vector with a positive zeroth component
to one with a negative zeroth component.
This guarantees the existence of a class of observer frames
for which a single root of the dispersion relation 
involves both positive and negative energies,
and it implies the canonical quantization procedure fails.

It is of interest to determine
the scale $\tilde{M}$ of the 3-momentum
at which the 4-momentum turns spacelike.
As an example,
consider the special case of the theory \rf{lagr}
with only a nonzero timelike $b_{\mu}$.
In the observer frame with $b_{\mu}=(b_0,\vec{0})$,
and taking $b_0 \sim \cO (m^2/M_P)$ according to the discussion
of scales in section 2,
it follows that $\tilde{M} \gsim {\cO}(M_P)$.
This shows that instabilities arise
only for Planck-scale 4-momenta in any of the concordant frames.
The associated negative-energy problem in boosted frames
emerges only for observers undergoing Planck-scale boosts.
The concordant-frame quantization described in section 3
therefore maintains stability 
for all experimentally attainable physical momenta
and in all experimentally attainable observer frames.

The existence of observer Lorentz invariance
implies that the negative-energy instabilities in strongly boosted frames
must have a counterpart in concordant frames,
albeit restricted to particles with Planck-scale energies.
This is indeed the case:
single-particle states with Planck-scale energies
in a concordant frame
are unstable to decay.
For example,
a Planck-energy fermion can explicitly be shown to be
unstable to the emission of a fermion-antifermion pair
\cite{kl}.
In conventional QED,
this process is kinematically forbidden.
However,
the presence of spacelike momenta
in the context of the Lorentz- and CPT-violating QED extension
makes the emission proceed at Planck energies.
Other conventionally forbidden processes are also likely to occur.
A single-particle state describing a fermion
of sufficiently large 3-momentum is therefore unstable
even in a concordant frame.

\it Causality. \rm
Microcausality holds in a quantum field theory
provided any two local observables with spacelike separation commute.
In a theory of Dirac fermions,
the local quantum observables are fermion bilinears.
Microcausality therefore holds for the modified theory \rf{lagr} if
\beq
iS(x-x^{\prime})=\{\psi(x),
\overline{\psi}(x^{\prime})\}=0,
\quad (x-x^{\prime})^2<0
\label{anticom}
\eeq
is satisfied.
Note that the original field $\ps$ is involved,
rather than the redefined field $\ch$,
because the definition of the latter 
depends on the choice of inertial frame.

An integral representation for $S(x-x^\prime)$
provides a useful tool in studying the conditions for microcausality.
In a concordant frame,
an integral represention for 
the anticommutator function $S(x-x^\prime)$
can be obtained as 
\cite{kl}
\beq
S(z)=
{\rm cof}(\Ga^{\mu} i\prt_{\mu}-M)
\int_{C} \fr{d^4\la} {(2\pi)^4}
\fr{e^{-i\la\cdot z}} {\det(\Ga^{\mu}\la_{\mu}-M)} .
\label{ffgreen}
\eeq
Provided
$ c_{\mu \nu} = d_{\mu \nu} = e_\mu = f_\mu = g_{\la \mu \nu}=0$,
the derivative couplings take
the standard form with $\Ga^{\mu}=\ga^{\mu}$.
In this case, 
a hermitian hamiltonian always exists
and the four poles remain on the real axis.
Explicit calculation of the contour integration
in Eq.\ \rf{ffgreen}
shows that $S(z)$ vanishes outside the light cone 
in this case.
It follows that the theory \rf{lagr} 
restricted to nonzero
$a_{\mu}$, $b_{\mu}$ and $H_{\mu \nu}$
is microcausal.
However, 
in the unrestricted theory \rf{lagr},
the poles of the integrand in Eq.\ \rf{ffgreen}
can lie away from the real $\la^0$ axis,
in which case the contour $C$ may fail to encircle them. 
This difficulty occurs when the bound on $\de^0$
discussed in section 3 is violated.
The hamiltonian then cannot be made hermitian,
and the roots of the dispersion relation may be complex.

In discussions of causality,
it is advantageous to define the velocity
of a particle at arbitrary 3-momentum.
However,
the definition of the quantum velocity operator 
is nontrivial even for the Lorentz- and CPT-invariant case
and becomes involved when these symmetries are violated
\cite{ck}.
One useful concept is the group velocity,
which can be defined for a monochromatic wave 
in terms of the dispersion relation by
$\vec v_g = {\prt E}/{\prt\vec{p}}$,
as usual.
It can be shown that 
the flow velocities of the conserved momentum 
and the conserved charge for one-particle states 
agree with the group velocity.
Also,
explicit checks in special cases suggest that 
$\langle d{\vec{x}}/dt\rangle =\vec{v}_{g}$
in relativistic quantum mechanics,
and that the maximal group velocity attainable
has magnitude equal to the maximal signal speed obtained from $S(z)$.

The group velocity can be used to establish the scale 
$\tilde{M}$
of microcausality violation.
Setting the group velocity to 1 and solving for the 
magnitude $|\vec p|$ of the 3-momentum 
determines the scale $\tilde{M}$ through 
$\tilde{M} = |\vec p|$.
For example,
consider the theory \rf{lagr}
with only the term $e_{\mu}$ nonzero
\cite{kl}.
Suppose $e_{\mu}$ is timelike,
and take $\vec{e}=0$ in the chosen concordant frame.
Then,
assuming the maximal expected Lorentz violation
$e_{0} \sim \cO(m/M_P)$
following the discussion in section 2,
the scale $\tilde M$ of microcausality violation
is obtained as $\tilde{M} \gsim {\cO}(M_P)$.
Thus,
the $e_{\mu}$ model violates microcausality 
at the scale $M_P$.

Intuition about
the connection between hermiticity 
of the hamiltonian $H$ and microcausality
can also be obtained
from the theory involving only $e_\mu$.
In this case,
the nonzero entries of the matrix $\ga^0\Ga^0$ 
in the Pauli-Dirac representation
consist only of diagonal entries $1 \pm e_0$.
For $|e_0|<1$, 
the spectrum of $\ga^0\Ga^0$ is therefore positive,
and both a suitable matrix $A$ 
and a hermitian hamiltonian $H$ exist.
However, 
when $|e_0|>1$
the spectrum of $\ga^0\Ga^0$ 
includes two negative eigenvalues,
and neither $A$ nor a hermitian $H$ exists.
In the dispersion relation,
the associated difficulty is that for $|e_0|>1$
it is always possible to find an observer frame
in which the roots become complex.

\it Intermediate-scale physics. \rm
The above results indicate that stability and causality
violations emerge at a scale ${\cO}(M_P)$.
However,
this conclusion may fail
for the special case of theories of the form \rf{lagr}
with a nonzero coefficient $c_{\mu\nu}$
\cite{kl}.
The point is that field operators with 
closely related derivative and spinor structures
are involved for both the usual Dirac kinetic term 
and the term with coefficient
$c_{\mu\nu}$,
which means the latter 
behaves in many respects as a first-order correction
to a zeroth-order result.
This feature is unique to the $c_{\mu\nu}$ term.

To illustrate this point with a definite example,
consider the special case of the lagrangian \rf{lagr}
with only the coefficient $c_{00}$ nonzero 
in a concordant frame,
and suppose in accordance with the discussion 
in section 2 that $c_{00} \sim \cO(m/M_P)$.
The dispersion relation for this model in an arbitrary frame is
\beq
(\et_{\al\mu}+c_{\al\mu})
(\et_{\pt{\nu}\nu}^{\al}+c^\al_{\pt{\nu}\nu})
\la^\mu\la^\nu
-m^2 = 0.
\label{cdispcov}
\eeq    
For the case $c_{00}>0$
it can then be shown that spacelike 4-momenta occur at a scale 
$\tilde{M} \gsim {\cO}(\sqrt{mM_P})$,
and so instabilities occur at energies well below
the scale $M_P$ of the underlying theory in this case.
If instead $c_{00}<0$,
then it can be shown that at the same scale
microcausality violations arise instead: 
the integration in (\ref{ffgreen})
can be performed analytically,
revealing that the anticommutator function $S(z)$ can be nonzero
outside the region defined by $z^0<(1+c_{00})|\vec{z}|$
and that signal propagation could therefore occur with maximal speed
$1/(1+c_{00})>1$.

These results concerning intermediate scales are of interest 
because energies comparable to the order of $\sqrt{mM_P}$ 
in the concordant frame are exhibited by certain physical phenomena.
For instance,
it has been suggested that effects from $c_{00}$-type terms 
might be responsible for the apparent excess of
cosmic rays in the region of $10^{19}$ GeV
\cite{cg,bc}. 
The above analysis suggests that these effects
can be traced to stability or causality violations,
which surely must be absent in a satisfactory underlying theory.
It therefore seems plausible that the effective dispersion relation 
involving the coefficients $c_{\mu\nu}$ is modified 
already at these scales,
perhaps along the lines described in the next section,
but in any case in a way that preserves stability and causality.
A corresponding modification of the predictions for cosmic rays
and other phenomena at the scale $\sqrt{mM_P}$ would then be likely.

\vglue 0.6 cm
\noindent
{\bf 5.\ PLANCK-SCALE EFFECTS}
\vglue 0.4 cm

The results described in section 4 indicate that problems 
with stability and causality in the theory \rf{lagr} 
arise primarily for Planck-scale 4-momenta in a concordant frame
or for observers undergoing a Planck boost relative to this frame.
Although perhaps strictly unnecessary from the phenomenological
viewpoint of current laboratory physics,
it would be of interest to establish
a framework for Lorentz and CPT violation
in which both stability and microcausality hold. 

A natural question is whether spontaneous Lorentz and CPT breaking
suffices to avoid the issues with stability and causality.
Since this type of breaking
could occur in a theory with a Lorentz-invariant lagrangian 
and hence with Lorentz-covariant dynamics,
it is likely to avoid at least some 
of the problems faced by theories with explicit
Lorentz and CPT violation.
For example,
one important advantage of spontaneous Lorentz violation
is the natural occurrence of observer Lorentz invariance,
which eliminates the coordinate dependence and related problems
faced by theories with explicit violation.
However,
spontaneous Lorentz violation manifests itself physically
because the Fock-space states are constructed on a noninvariant vacuum,
whereas stability and causality in a relativistic theory 
depend partly on the existence of a Lorentz-invariant vacuum.
It is therefore to be expected that,
despite the advantages of spontaneous Lorentz violation,
some difficulties with stability and causality remain.

This expectation can be directly confirmed by studying 
the fermion sector of quantum field theories 
with spontaneous Lorentz violation 
\cite{kl}. 
The key point is that the lagrangian \rf{lagr}
includes by construction
the most general terms quadratic in fermion fields 
that appear in a renormalizable theory.
This means that implementing
spontaneous Lorentz and CPT violation
in any conventional fermion field theory 
must result in free-fermion Fock-space states
with dispersion relations described by Eq.\ \rf{disp}
or a restriction of it.
If indeed all possible dispersion relations have 
either stability or causality violations at some large scale,
then no fermion lagrangian 
with spontaneous Lorentz and CPT violation
can have a completely satisfactory perturbative Hilbert space
in conventional quantum field theory. 
Maintaining stability and causality would therefore require
an additional ingredient beyond conventional quantum field theory,
in accordance with the notion that
Lorentz and CPT violation is a unique potential signal 
for Planck-scale physics.

The above discussion suggests that a theory
with a quadratic lagrangian
that maintains both stability and causality
would need to include terms beyond the ones in Eq.\ \rf{lagr}.
In the low-energy limit 
of an underlying realistic theory at the Planck scale
with spontaneous Lorentz violation,
any such terms would emerge 
as higher-dimensional nonrenormalizable operators 
that must be included in the standard-model extension 
at energies determined by the Planck scale.
The structure of the dispersion relation 
would correspondingly change.
In a concordant frame,
it would need to remain of the form \rf{disp} for small 3-momenta
but would avoid spacelike 4-momenta and group velocities exceeding 1
for large 3-momenta.
In fact,
it can be shown by explicit construction
that appropriate modifications of the dispersion relation  
can satisfy both the stability and causality requirements
\cite{kl}.
It suffices to introduce
a suitable factor suppressing the coefficients 
for Lorentz and CPT violation only at large 3-momenta.
Since the notion of large 3-momenta is frame dependent,
the suppression factor itself must be frame-dependent
and hence must involve Lorentz- and CPT-violating coefficients.
 
As an example,
consider the dispersion relation \rf{disp} 
in the special case with only $c_{00}$ nonzero and negative
in a concordant frame.
In an arbitrary frame,
this dispersion relation has the form \rf{cdispcov}.
To minimize complications arising from 
the small size of the coefficients for Lorentz and CPT violation,
let the mass and the value of the Lorentz-violating coefficients 
be of order 1 in appropriate units. 
If each factor of $c_{00}$ is combined with
an exponential factor $\exp(c_{00} \la_0^2)$,
then in an arbitrary frame
the dispersion relation \rf{cdispcov} becomes 
\beq
(\et_{\al\mu}+c_{\al\mu} \exp(c_{\be\ga}\la^\be\la^\ga))
(\et^\al_{\pt{\nu}\nu}+c^\al_{\pt{\nu}\nu} \exp(c_{\be\ga}\la^\be\la^\ga))
\la^\mu\la^\nu
-m^2 = 0.
\label{cdispmod}
\eeq
The presence of the exponential factors eliminates 
the violations of microcausality 
that occurred at large $\la_\mu$ in Eq.\ \rf{cdispcov}.
The group velocity now lies below 1 for all $\vec \la$.
It can also be shown that no stability problems are introduced
\cite{kl}.

It would evidently be interesting to identify theories 
in which dispersion relations of this type naturally arise.
If transcendental functions of the momenta
are indeed necessary to overcome the polynomial 
Lorentz-violating behavior in Eq.\ \rf{disp},
a suitable lagrangian must incorporate
derivative couplings of arbitrary order.
It then follows that 
spontaneous Lorentz and CPT violation in a nonlocal theory 
can naturally provide the required structure 
for stability and causality at all scales.

The emergence of nonlocality as an important ingredient
is noteworthy in part because string theories 
are nonlocal objects.
Moreover,
calculations with string field theory 
provided the original motivation
for identifying spontaneous Lorentz and CPT violation 
as a potential Planck-scale signal 
\cite{kps}
and for the development of the standard-model extension
as the relevant low-energy limit
\cite{kp,ck}.
In fact,
it can be shown that the 
structure of the field theory for the open bosonic string
is compatible with dispersion relations of the desired type
\cite{kl}.
Thus,
suppose in this string theory there exist
nonzero Lorentz- and CPT-violating expectation values of tensor fields
resulting from spontaneous symmetry breaking.
Then,
for example,
the dispersion relation for the scalar tachyon mode
contains a piece closely related to Eq.\ \rf{cdispcov},
but it includes also nonlocal terms of the type 
needed to maintain stability and causality.
Although the bosonic string is merely a toy model in this context,
the structural features of interest are generic to other string theories,
including ones with fermions.
This provides support for the existence of 
a stable and causal realistic fundamental theory
exhibiting spontaneous Lorentz violation.

\vglue 0.6 cm
\noindent
{\bf 6.\ ACKNOWLEDGMENTS}
\vglue 0.4 cm

This work is supported in part
by the United States Department of Energy 
under grant number DE-FG02-91ER40661.

\newpage
\noindent
{\bf 7.\ REFERENCES}
\vglue 0.4 cm

\end{document}